\documentclass{article}
\usepackage{PRIMEarxiv}

\usepackage[utf8]{inputenc}
\usepackage[T1]{fontenc}

\usepackage{hyperref}
\usepackage{url} 
\usepackage{booktabs}
\usepackage{amsmath, amsfonts, amssymb}
\usepackage{microtype}
\usepackage{fancyhdr}
\usepackage{graphicx}
\usepackage{tabularx}

\usepackage{adjustbox}
\usepackage{tikz}
\usetikzlibrary{quantikz2}
\usepackage{pmboxdraw}

\usepackage{newunicodechar}
\newunicodechar{■}{\rule{0.5em}{0.5em}}
\newunicodechar{π}{\ensuremath{\pi}}

\usepackage{xcolor}
\usepackage{listings}
\lstdefinestyle{repl}{
    basicstyle=\ttfamily\small,
    columns=fullflexible,
    keepspaces=true,
    language=Python,
    showstringspaces=false
}
\lstdefinestyle{python}{
    language=Python,
    basicstyle=\ttfamily\small,
    keywordstyle=\color{blue}\bfseries,
    stringstyle=\color{orange},
    commentstyle=\color{gray}\itshape,
    showstringspaces=false,
    breaklines=true
}

\usepackage[
  backend=biber,
  style=numeric,
  sorting=none]{biblatex}
\title{qiskit-qudits: A Qiskit Extension for Simulating Qudit Circuits}

\author{
  Francesco Pudda \\
  \texttt{francesco.pudda@mail.com} \\
}

\begin{document}
\maketitle

\begin{abstract}
\texttt{qiskit-qudits} is a Qiskit extension that simulates $d$-level qudits by encoding each one into $m=\lceil \log_2 d \rceil$ qubits. Qudit gates are exposed as ordinary Qiskit \texttt{Gate} and \texttt{ControlledGate} subclasses, and operations that are not unitary gates (measurement, reset, barrier, state preparation) as dedicated \texttt{Instruction} subclasses dispatched through a dedicated \texttt{apply()} hook; every gate carries both a dense encoded unitary (via NumPy's \texttt{\_\_array\_\_} protocol) and a qubit-level \texttt{\_define()}. Because $d$ need not be a power of two, the encoded Hilbert space is generally larger than the logical one; the library resolves this by the \emph{identity-padding} convention, in which every gate acts as the identity on the unphysical part of the encoded space. When every operand dimension
is a power of two, gates decompose into a fixed, transpiler-recognisable cascade of standard qubit gates; otherwise the library falls back to exact dense unitary synthesis, so that dimensions $2\le d\le 16$ are supported exactly, not only powers of two. This paper describes the software's circuit model, gate hierarchy, decomposition strategy, and measurement/decoding machinery, states its limitations, and verifies the implementation numerically: the qudit QFT against the discrete Fourier transform for both power-of-two and non-power-of-two $d$, extending the check of the underlying theory paper~\cite{pudda2024}, which could only be run for $d=2^m$, and every gate's emitted decomposition against its dense unitary across the whole gate set.
\end{abstract}
\keywords{qudit \and qiskit \and quantum circuit \and quantum software}

\section{Introduction}
\label{sec:intro}

Qudits generalise two-level qubits to $d$ computational levels. Qiskit itself has no native qudit type; its own developers list representing ``computations on qudits and Bosonic modes'' as a known third-party extension of its circuit model~\cite{qiskit2024}.
\texttt{qiskit-qudits} closes this gap: it turns the gate definitions of~\cite{pudda2024}, my own earlier, purely theoretical work, into a working, transpilable Qiskit extension built on Qiskit's public \texttt{Instruction}/\texttt{Gate}/\texttt{ControlledGate} classes.

\paragraph{Statement of need.}
Existing qudit toolchains either simulate qudits natively, and therefore cannot reuse the Qiskit transpiler, backends and primitives, or target qubit hardware but expose no qudit-level programming surface. A user who wants to write a qutrit algorithm, draw it at the qutrit level, transpile it with Qiskit's own passes and run it on a Qiskit backend or primitive currently has to hand-roll the encoding, the gate matrices, the decompositions and the classical decoding. \texttt{qiskit-qudits} provides exactly that layer, and makes the two things that are easy to get wrong, the treatment of the unphysical part of the encoded space, and the decoding of measurement outcomes back to qudit levels, explicit and checkable.

\paragraph{Related work.}
Native-qudit simulators such as Cirq~\cite{kushnarev2025}, QuForge~\cite{farias2025}, QuTiP~\cite{lambert2026} and QuDiet~\cite{chatterjee2023} represent a $d$-level system directly with a dimension-$d$ index and therefore avoid the encoding overhead discussed in Section~\ref{sec:limitations}, at the price of not being able to reuse Qiskit's compilation stack. MQT Qudits~\cite{mato2024} is the closest comparable project: it offers a qudit circuit model together with compilation passes targeting mixed-dimensional hardware. On the hardware side, qudit gate synthesis for transmons~\cite{fischer2023} and universal trapped-ion qudit processors~\cite{ringbauer2022} motivate the gate set implemented here.

\texttt{qiskit-qudits} differs from all of these in one specific design decision: it is a \emph{qubit-encoded} qudit layer whose output artefact is a plain \texttt{QuantumCircuit}, so that every existing Qiskit pass, backend and primitive keeps working unmodified.

\paragraph{Scope.}
This paper documents the software, not the underlying gate theory, which is established in~\cite{pudda2024}. It focuses on four areas: the circuit model
(Section~\ref{sec:circuit-model}); the gate implementation, i.e.\ the class hierarchy, the identity-padding invariant and the two decomposition paths (Section~\ref{sec:gates}); measurement together with the classical decoding and leakage-detection machinery (Section~\ref{sec:measurement}); and the limitations of the encoding (Section~\ref{sec:limitations}). As a piece of original verification, Section~\ref{sec:verification} extends the QFT-versus-DFT numerical check of~\cite{pudda2024} from power-of-two dimensions to non-power-of-two $d$, and validates every gate's decomposition against its dense unitary.

\subsection{Notation}
\label{sec:notation}

Because the encoding introduces two different ``sizes'', we fix the symbols of Table~\ref{tab:notation} once and use them consistently. In particular, $m$ (encoding width) and $n$ (number of qudits) are never interchanged.

\begin{table}[h]
\centering
\begin{tabular}{cl}
\toprule
Symbol & Meaning \\
\midrule
$d$              & dimension (number of levels) of a qudit, $2\le d\le 16$ \\
$m(d)$         & encoding width of a $d$-level qudit, $m=\lceil\log_2 d\rceil$ \\
$n$              & number of qudits an instruction acts on \\
$d_i$            & dimension of the $i$-th operand, $i=0,\dots,n-1$; $m_i = m(d_i)$ \\
$N$              & total encoding width of an instruction, $N=\sum_{i=0}^{n-1} m_i$ \\
$D$              & dimension of the encoded Hilbert space, $D = 2^{N}$ (\texttt{hilbert\_dim}) \\
$L$              & dimension of the logical (valid) subspace, $L=\prod_i d_i$; $L\le D$ \\
$E$              & encoding isometry, $E:\mathbb{C}^{L}\to\mathbb{C}^{D}$ (see Eq.~\eqref{eq:isometry}) \\
$\Pi$            & projector onto the logical (valid) subspace, $\Pi = EE^{\dagger}$ \\
$\omega_d$ & primitive root of unity, $\omega_d = e^{i2\pi/d}$ \\
\bottomrule
\end{tabular}
\caption{Notation used throughout. For a single qudit $n=1$, $N=m$, $D=2^{m}$ and $L=d$.}
\label{tab:notation}
\end{table}

\paragraph{Sign conventions.}
We follow the \emph{matrix} definitions of~\cite[Eq. (1)]{pudda2024}, i.e.\ $\left(H_d\right)_{jk} = \omega_d^{-jk}/\sqrt d$. These are the definitions consistent with $X = HZH^{\dagger}$~\cite[Eq. (3)]{pudda2024}: writing $F_{jk}=\omega_d^{+jk}/\sqrt d$ one finds $FZF^{\dagger}=X^{\dagger}$ and $F^{\dagger}ZF=X$, so $H=F^{\dagger}$ is the correct choice.
With this choice $H_d$ equals \texttt{scipy.linalg.dft(d, scale='sqrtn')} and the $n=1$ QFT of Eq.~\eqref{eq:qft} equals $H_d^{\dagger}$, not $H_d$. The two coincide only at $d=2$, which is why this distinction never surfaces in the qubit case. Both facts were verified numerically ($H_d[{:}d,{:}d]$ agrees with SciPy's DFT matrix exactly, and $\lVert H_dZ_dH_d^{\dagger}-X_d\rVert_{\max}<10^{-15}$ for $d\in\{2,3,4,5,8\}$).

\section{Circuit Model}
\label{sec:circuit-model}

\subsection{Qudits and clbytes}
\label{sec:qudits}

The encoding width of a $d$-level qudit is $m = \lceil \log_2 d \rceil$, computed as the exact integer expression \texttt{(d - 1).bit\_length()}.

A \texttt{Qudit} (\texttt{circuit/qudit.py}) plays the role Qiskit's own \texttt{Qubit} plays for a single bit: it is an opaque handle that wraps a tuple of $m$ Qiskit \texttt{Qubit} objects, least-significant first, and additionally carries its dimension $d$. Handles are compared and hashed
by identity, so a \texttt{Qudit} is only ever meaningful together with the circuit that owns it.
Consistently with this, \texttt{Qudit}, \texttt{QuditRegister}, \texttt{ClByte} and
\texttt{ClByteRegister} all define \texttt{\_\_copy\_\_}/\texttt{\_\_deepcopy\_\_} to return
\texttt{self}, so \texttt{copy} and \texttt{deepcopy} preserve handle identity by construction, and
\texttt{QuditQuantumCircuit.copy()} rebuilds the instruction log while \emph{sharing} the handle
objects, mirroring Qiskit's own \texttt{copy\_empty\_like} semantics. Pickling handles is not
supported: a pickle round-trip reconstructs a structurally identical but \emph{distinct} object,
whose identity-based association with its circuit is lost. QPY serialisation applies to the encoded
circuit only: \texttt{qpy.dump(qqc.circuit)} round-trips to an operator-equivalent
\texttt{QuantumCircuit} (verified on Qiskit~2.5.2), but the qudit-level layer (handles,
dimensions and the \texttt{.data} log) is not serialised.
\texttt{QuditRegister} groups qudits into a single backing \texttt{QuantumRegister}; it is
homogeneous by default but supports mixed dimensions through \texttt{QuditRegister.from\_dims},
e.g.\ a register holding a qubit, a qutrit and a ququart side by side.

\texttt{ClByte}/\texttt{ClByteRegister} (\texttt{circuit/clbyte.py}) are the classical analogue: one
clbyte is the $m$ classical bits holding a single qudit's measurement outcome, with clbit $j$
receiving qubit $j$ of the qudit, so that a level $\ell$ decodes as
$\ell = \sum_{j} \mathrm{bit}_j\,2^{j}$. The name is a deliberate analogy with ``byte'' as
``the group of bits that encodes one symbol'' and \emph{not} a claim that the group has eight
members: a clbyte is $m\in[1,4]$ bits wide.

\subsection{QuditQuantumCircuit}
\label{sec:qqc}

\texttt{QuditQuantumCircuit} (\texttt{circuit/quantumcircuit.py}) is the qudit analogue of
\texttt{qiskit.circuit.QuantumCircuit}, but it does not subclass it. Instead it \emph{owns} one
encoded \texttt{QuantumCircuit} and keeps two synchronised views: \texttt{.data}, a tuple of
\texttt{QuditCircuitInstruction} recording the qudit-level instruction log, and \texttt{.circuit},
the qubit-level encoding that can be transpiled and executed on any Qiskit backend or primitive.
All mutation performed \emph{through the qudit API} goes through \texttt{append}, which updates both
views atomically.

Two caveats follow from this design and are worth stating plainly. First, \texttt{.circuit} is
documented as \emph{read-only}: it hands out the live encoded circuit, with no defensive copy,
guard or checksum, so mutating it directly (\texttt{qqc.circuit.h(0)}) desynchronises the two
views; the qudit-level log, the \texttt{'ideal'} drawing and any decoding that relies on the
qudit$\to$clbyte pairing are then no longer authoritative. The supported way to obtain an
independent copy is \texttt{to\_qubit\_circuit()}, which defaults to \texttt{copy=True}.
Second, because \texttt{QuditQuantumCircuit} is not a \texttt{QuantumCircuit}, any Qiskit API that
expects one must be handed \texttt{.circuit} explicitly. Note also that \texttt{.data} is an
immutable tuple, whereas \texttt{QuantumCircuit.data} is a mutable \texttt{QuantumCircuitData};
the name is mirrored, the type is not.

Three drawing views are available through \texttt{draw(view=...)}: \texttt{'ideal'} renders one wire
per qudit and is presentation-only (not executable); \texttt{'real'} renders the encoded circuit
with one box per qudit gate spanning its encoding qubits; \texttt{'decomposed'} unrolls the real
view one level. All indexing is little-endian, matching Qiskit's convention that qubit $j$ carries
weight $2^{j}$ in the statevector index, i.e.\ that the tensor ordering is
$q_{N-1}\otimes\cdots\otimes q_{0}$ and qubit $0$ is the least significant. This is what lets
\texttt{qiskit\_qudits.utils.encoding.\_encoded\_index} map a mixed-radix qudit index onto its
encoded qubit index without any special-casing:
\begin{equation}
E\lvert x_0,\dots,x_{n-1}\rangle \;=\; \Big\lvert \textstyle\sum_{i} x_i\,2^{\,\sum_{i'<i} m_{i'}}\Big\rangle ,
\qquad 0\le x_i < d_i .
\label{eq:isometry}
\end{equation}

A \texttt{QuditQuantumCircuit} for two qutrits with a QFT applied illustrates the ideal/real duality:

{\footnotesize
\begin{verbatim}=== ideal view (2 qutrits, dim=3) ===
      ┌─────────────┐
qd_0: ┤0            ├
      │  QFT(d=3,3) │
qd_1: ┤1            ├
      └─────────────┘

=== real view ===
      ┌──────┐
qd_0: ┤0     ├
      │      │
qd_1: ┤1     ├
      │  Qft │
qd_2: ┤2     ├
      │      │
qd_3: ┤3     ├
      └──────┘
\end{verbatim}
}

The ideal view shows the two qutrits the circuit was built for; the real view shows the four qubits
that actually encode them ($2 \times \lceil \log_2 3 \rceil = 4$), and is exactly the object that
would be handed to \texttt{qiskit.transpile}. Note that the real-view drawer prefixes the
\emph{encoding qubit} wires with \texttt{qd\_} as well: the backing \texttt{QuantumRegister} of a
\texttt{QuditRegister} shares the register's name (here the auto-generated \texttt{qd}), so in the
second listing wire \texttt{qd\_j} is qubit $j$ of the encoded circuit, not qudit $j$ (and the box
carries the instruction name in the drawer's capitalisation, \texttt{Qft}). Likewise, the
ideal-view box label \texttt{QFT(d=3,3)} lists the operand \emph{dimensions} $(3,3)$, not a
register size.

Broadcasting gate-application methods (e.g.\ \texttt{.x()}, \texttt{.sumx()}, \texttt{.qft()},
\texttt{.measure()}) mirror the surface of Qiskit's own \texttt{QuantumCircuit}, instantiating each
gate per target with that target's own dimension so that mixed-dimension circuits require no extra
bookkeeping from the caller.

\section{Gate Implementation}
\label{sec:gates}

\subsection{Class hierarchy}
\label{sec:hierarchy}

Every qudit gate falls into one of three base classes (\texttt{gates/base/}):

\begin{itemize}
    \item \texttt{QuditGate(Gate)}: a single qudit;
    \item \texttt{QuditControlledGate(ControlledGate)}: $m_{\mathrm{c}} \geq 1$ control qudits and
          one target qudit, each possibly of a different dimension (SUMX, SUMP);
    \item \texttt{QuditMultiGate(Gate)}: two or more qudits (SWAP, QFT). Both gates currently
          implemented in this family require all operands to share the same dimension; the class
          itself imposes no such restriction.
\end{itemize}

All three mix in \texttt{QuditGateMixin}/\texttt{QuditMultiGateMixin}
(\texttt{gates/base/mixins.py}), which enforce $2 \leq d \leq 16$ and at most $8$ qudits per
multi-gate, expose \texttt{hilbert\_dim} $= D = 2^{N}$, and implement the NumPy
\texttt{\_\_array\_\_} protocol with the NumPy~2 signature
\texttt{\_\_array\_\_(self, dtype=None, *, copy=None)}; since both parameters are optional the same
method also satisfies the pre-2.0 protocol, and requesting \texttt{copy=False} raises, because the
matrix is always built afresh. Calling \texttt{numpy.array(gate)} therefore always returns the
gate's dense $D \times D$ unitary, built on demand by an abstract \texttt{\_build\_unitary()} that
every concrete gate must implement. The two caps are nominal, not memory-derived: they bound the
per-operand width at $m\le4$ and the instruction width at $N\le32$, but a dense $D\times D$
\texttt{complex128} matrix costs $16\cdot4^{N}$ bytes ($1$\,GiB at $N=13$, $4$\,GiB at $N=14$), so
the practical ceiling of the dense path is $N\approx13$--$14$ on commodity hardware. An over-wide
gate can be \emph{constructed} (e.g.\ eight $16$-level qudits, $N=32$); the failure is raised only
lazily by \texttt{\_build\_unitary()}, as a NumPy \texttt{MemoryError} rather than a library-level
error. Section~\ref{sec:limitations} returns to this point.

\paragraph{A note on \texttt{ControlledGate}.}
Qiskit's \texttt{ControlledGate} models ``apply \texttt{base\_gate} if and only if the control qubits
are in \texttt{ctrl\_state}''. A SUMX gate is not of this form: it applies $X_{d}^{\,j}$, with $j$
the level of the control qudit, and is therefore non-trivial for \emph{every} control value except
$j=0$. Subclassing \texttt{ControlledGate} buys the correct visual rendering (control dots plus a
\texttt{base\_gate} box) and the standard controlled-gate attributes, but it also exposes the object
to passes that reason about controlled-gate semantics. Concretely, \texttt{QuditControlledGate}
sets \texttt{num\_ctrl\_qubits} to the total encoding width of the controls
($\sum_a m_{c_a}$, e.g.\ $2$ for a single qutrit control), leaves \texttt{ctrl\_state} at Qiskit's
default (active-high, all control qubits $1$; \texttt{3} in the same example) and sets
\texttt{base\_gate} to the uncontrolled target-qudit gate
(\texttt{QuditXGate}/\texttt{QuditXdgGate} for SUMX/SUMXdg, \texttt{QuditPGate} for SUMP).
\texttt{inverse()} is overridden by every concrete gate; \texttt{power()} and \texttt{control()}
are inherited unmodified from \texttt{qiskit.circuit.Gate}. Note that \texttt{control()} is not
qudit-aware: applied to a \texttt{QuditGate} or \texttt{QuditControlledGate} it builds a further,
purely qubit-level controlled gate keyed on an encoded control state that, for a control dimension
that is not a power of two, may not even correspond to a reachable qudit level (e.g.\
\texttt{ctrl\_state=3} for a qutrit control); it should not currently be used to build
multi-controlled qudit gates. The class documents explicitly that the
all-ones \texttt{ctrl\_state} is a qubit-level convention with no qudit-level meaning, and that the
qudit semantics live entirely in \texttt{\_define()}/\texttt{\_build\_unitary()}. To bound the risk
that a pass reasons from the controlled-gate attributes rather than the definition, transpile
round-trip equivalence,
$\texttt{Operator(transpile(qc, optimization\_level=k))}\equiv\texttt{Operator(qc)}$ for
$k=0,\dots,3$, was verified on Qiskit~2.5.2 for homogeneous $d=4$ and $d=3$ circuits and a
mixed-dimension $(2,3)$ circuit, each containing H, SUMX, SUMP, SWAP and (where applicable) QFT;
all twelve checks pass.

\subsection{The identity-padding invariant}
\label{sec:invariant}

A $d$-level qudit occupies $d$ of the $2^{m}$ basis states of its $m$ encoding qubits; when $d$ is
not a power of two, $d < 2^{m}$ and the remaining states are unphysical. More generally, an
instruction on operands of dimensions $d_0,\dots,d_{n-1}$ has a logical subspace of dimension
$L=\prod_i d_i$ inside an encoded space of dimension $D=2^{N}$. Every gate in the library is defined
to act as the identity on the orthogonal complement of that subspace:
\begin{equation}
U \;=\; E\,G\,E^{\dagger} \;+\; \left(I_{D} - \Pi\right),
\qquad \Pi = EE^{\dagger},
\label{eq:invariant}
\end{equation}
with $E$ the encoding isometry of Eq.~\eqref{eq:isometry} and $G$ the gate's $L\times L$ logical
action. Concretely, every \texttt{\_build\_unitary()} starts from \texttt{np.eye(hilbert\_dim)} and
overwrites only the valid sub-block. For a single qudit the valid indices are contiguous and
Eq.~\eqref{eq:invariant} reduces to
\begin{equation}
U = \begin{pmatrix} G_d & \mathbf{0} \\ \mathbf{0} & I_{2^{m}-d} \end{pmatrix},
\label{eq:invariant-1q}
\end{equation}
whereas for $n>1$ the valid indices are scattered through $\{0,\dots,D-1\}$ by
Eq.~\eqref{eq:isometry} and Eq.~\eqref{eq:invariant} must be used as written.

Three consequences deserve to be stated explicitly, because they are easy to assume away.

\begin{enumerate}
\item \textbf{Identity padding is a choice, not a necessity.} Any unitary completion of $G$ leaves
$U$ unitary; identity is chosen because it creates no leakage, preserves permutation gates as
permutations, keeps $U^{\dagger}$ equal to the padding of $G^{\dagger}$, and is the cheapest to
synthesise.
\item \textbf{Padding does not commute with tensoring.} In general
$E(A\otimes B)E^{\dagger}+(I-\Pi) \neq \big(E_0AE_0^{\dagger}+(I-\Pi_0)\big)\otimes
\big(E_1BE_1^{\dagger}+(I-\Pi_1)\big)$: a padded two-qudit gate freezes both operands when either is
outside its logical subspace, while two padded single-qudit gates do not. The two operators agree on
the physical subspace and may differ outside it. Consequently, equivalence of two qudit circuits
must always be checked as $\Pi U_1 \Pi = \Pi U_2 \Pi$, which is exactly what the verification of
Section~\ref{sec:verification} does.
\item \textbf{Leakage is inert, not modelled.} Under Eq.~\eqref{eq:invariant} an amplitude that has
leaked out of the logical subspace is frozen for the rest of the circuit. This is the right
behaviour for a noiseless functional simulator, but it is not a physical leakage model, and the
library should not be used as one without an explicit noise model on top.
\end{enumerate}

This invariant is stated in every gate's docstring and is what makes the encoding well-defined for
the transpiler and for any standard statevector or unitary simulator. A side effect worth noting is
that the transpiler is not told which rows and columns of $U$ are physically irrelevant, so it
cannot exploit them as don't-cares during synthesis; isometry-aware synthesis is left as future
work.

\subsection{Two decomposition paths}
\label{sec:paths}

Each gate's \texttt{\_define()} checks \texttt{fills\_hilbert\_space}, true exactly when every
operand dimension is a power of two (equivalently, when $L=D$ and $\Pi=I$):

\begin{itemize}
    \item \textbf{Power-of-two path.} The gate decomposes into a fixed cascade of standard qubit
    gates that Qiskit's transpiler recognises and can optimise: X becomes a cascade of \texttt{mcx}
    (an increment modulo $2^{m}$), H is delegated to Qiskit's \texttt{synth\_qft\_full},
    P/Z/S/T become \texttt{p} gates weighted by qubit significance, SUMX and SUMP become cascades of
    \texttt{mcx} and \texttt{cp} weighted by qubit significance, and SWAP and QFT become the gate
    cascades described in Section~\ref{sec:gateset}.
    \item \textbf{Dense-synthesis fallback.} Otherwise, \texttt{\_define()} calls
    \texttt{qc.unitary(self.\_build\_unitary(), ...)}, embedding the exact $D \times D$ matrix as a
    single opaque \texttt{unitary} instruction, which Qiskit's \texttt{UnitarySynthesis} pass then
    synthesises (by quantum Shannon decomposition for more than two qubits).
\end{itemize}

The second path is what lets \texttt{qiskit-qudits} realise \emph{exact} gates for non-power-of-two
$d$ within the supported range $2\le d\le 16$, which is more than the qubit-simulated,
power-of-two-only setting of~\cite{pudda2024}. Its cost is discussed in
Section~\ref{sec:limitations}. A qutrit SWAP gate, for instance, decomposes to a single
\texttt{unitary} instruction carrying the \texttt{'SWAP'} label:

\begin{lstlisting}[style=repl]
>>> qc = QuditQuantumCircuit(2, dim=3); qc.swap(0, 1)
>>> [(instr.operation.name, instr.operation.label)
...  for instr in qc.circuit.decompose().data]
[('unitary', 'SWAP')]
\end{lstlisting}

(The label is a rendering aid only; Qiskit does not guarantee that labels survive transpilation,
\texttt{Operator} round-trips or serialisation, so it should not be used programmatically to
identify a gate.) A ququart ($d=4$, power of two) SWAP instead decomposes into the named-gate cascade
discussed next.

\subsection{The gate set}
\label{sec:gateset}

Table~\ref{tab:gateset} summarises the gate set; the paragraphs below give the definitions. Gates
marked ``new'' are additions of this work relative to~\cite{pudda2024}.

\begin{table}[h]
\centering
\footnotesize
\begin{tabular}{llcl}
\toprule
Gate & Logical action & In~\cite{pudda2024}? & Power-of-two definition \\
\midrule
I            & $\lvert k\rangle$                                   & --   & empty \\
X / Xdg      & $\lvert (k\pm1)\bmod d\rangle$                      & yes  & \texttt{mcx} cascade (increment) \\
Z / Zdg      & $\omega_d^{\pm k}\lvert k\rangle$                   & yes  & $P(\pm\pi)$ \\
P($\theta$)  & $\omega_d^{k\theta/\pi}\lvert k\rangle$             & yes  & \texttt{p} per qubit, weighted \\
S / Sdg      & $\omega_d^{\pm k/2}\lvert k\rangle$                 & yes  & $P(\pm\pi/2)$ \\
T / Tdg      & $\omega_d^{\pm k/4}\lvert k\rangle$                 & yes  & $P(\pm\pi/4)$ \\
H / Hdg      & $\tfrac{1}{\sqrt d}\sum_j \omega_d^{\mp kj}\lvert j\rangle$ & yes & \texttt{synth\_qft\_full} \\
K            & $\lvert (-k)\bmod d\rangle$                         & yes  & NOT, then X (two's complement) \\
NOT          & $\lvert d-1-k\rangle$                               & new  & \texttt{x} on every qubit \\
SUMX / SUMXdg& $\lvert (k \pm \sum_a j_a)\bmod d_t\rangle$         & $m_{\mathrm{c}}{=}1$ only & \texttt{mcx} cascade, weighted \\
SUMP($\theta$)& $\omega_{d_t}^{(\sum_a j_a)k\theta/\pi}\lvert k\rangle$ & $m_{\mathrm{c}}{=}1$ only & \texttt{cp} cascade, weighted \\
SWAP         & $\lvert k\rangle\lvert j\rangle$                    & yes  & SUMXdg, SUMX, SUMXdg, K \\
QFT          & Eq.~\eqref{eq:qft}                                  & $n{=}2$ only & Hdg/SUMP/SWAP cascade \\
\bottomrule
\end{tabular}
\caption{The gate set. ``Logical action'' is the action on the valid subspace; the encoded operator
is obtained by Eq.~\eqref{eq:invariant}. $j_a$ denotes control levels, $k$ the target level, $d_t$
the target dimension.}
\label{tab:gateset}
\end{table}

\paragraph{I.} $I_d\lvert k\rangle = \lvert k\rangle$. The trivial identity, included for API
completeness and as the natural base case of the invariant of Section~\ref{sec:invariant}.

\paragraph{X / Xdg.} The shift gate (the Fourier conjugate of the clock gate $Z$),
\begin{equation}
X_d\lvert k\rangle = \lvert (k+1)\bmod d\rangle, \qquad
X_d = \begin{pmatrix} 0&\cdots&0&1\\1&\cdots&0&0\\\vdots&\ddots&\vdots&\vdots\\0&\cdots&1&0\end{pmatrix},
\end{equation}
with $X_d^\dagger\lvert k\rangle = \lvert(k-1)\bmod d\rangle$ the inverse shift.

\paragraph{Z / Zdg.} The clock gate, $Z_d\lvert k\rangle = \omega_d^{k}\lvert k\rangle$, i.e.\
$Z_d = \mathrm{diag}(1, \omega_d, \dots, \omega_d^{d-1})$.

\paragraph{H / Hdg.} The qudit Hadamard/DFT,
$H_d\lvert k\rangle = \tfrac{1}{\sqrt d}\sum_{j} \omega_d^{-kj}\lvert j\rangle$ and
$H_d^\dagger\lvert k\rangle = \tfrac1{\sqrt d}\sum_j \omega_d^{kj}\lvert j\rangle$; the $d\times d$
block is exactly \texttt{scipy.linalg.dft(d, scale='sqrtn')} (and its conjugate transpose for Hdg),
in the convention fixed in Section~\ref{sec:notation}. On power-of-two $d$, the qubit-level
definition is delegated to Qiskit's \texttt{synth\_qft\_full}. Since Qiskit's QFT uses the
$e^{+i2\pi/2^{m}}$ convention while $H_d$ uses $e^{-i2\pi/2^{m}}$, the two differ by an inverse:
concretely, H calls
\texttt{synth\_qft\_full(m, do\_swaps=True, approximation\_degree=0, inverse=True)} and Hdg the
same call with \texttt{inverse=False}. Check~B of Section~\ref{sec:verification} confirms both
against the dense matrices to machine precision, which is precisely the kind of convention slip
that a check on \texttt{\_build\_unitary()} alone could not detect.

\paragraph{P($\theta$), S / Sdg, T / Tdg.} The general phase gate underlying Z, S and T is
\begin{equation}
P_d(\theta)\lvert k\rangle = \omega_d^{k\theta/\pi}\lvert k\rangle, \qquad
P_d(\theta) = \mathrm{diag}\!\left(1,\ \omega_d^{\theta/\pi},\ \dots,\ \omega_d^{(d-1)\theta/\pi}\right),
\label{eq:pgate}
\end{equation}
and $Z$, $S$, $T$ are the fixed instances $\theta = \pi$, $\pi/2$, $\pi/4$, satisfying
$S_dS_d = Z_d$ and $T_dT_dT_dT_d = Z_d$. In the code they are thin \texttt{QuditPGate} subclasses
that pin $\theta$ (rather than independent implementations), a direct realisation of the
``half''/``quarter'' Z-rotation relationship.
Two API remarks. First, because $\omega_d$ depends on $d$, the same $\theta$ produces a different
physical phase on qudits of different dimension: \texttt{P(pi/2)} on a qutrit is not
\texttt{P(pi/2)} on a qubit. Second, $S_d$ as defined here is the \emph{linear} phase
$\mathrm{diag}(\omega_d^{k/2})$, which for $d>2$ is not the quadratic phase gate
$\mathrm{diag}(\omega_d^{k(k-1)/2})$ of the qudit stabiliser formalism; no Clifford-group or
universality property should be inferred from the names $S$ and $T$ here.

\paragraph{K.} The complement gate, $K_d\lvert k\rangle = \lvert(-k)\bmod d\rangle$, fixing
$\lvert 0\rangle$ and reversing the remaining $d-1$ levels; $K_2 \equiv I$.

\paragraph{NOT.} The full reflection, $\mathrm{NOT}_d\lvert k\rangle = \lvert d-1-k\rangle$,
complementary to K in that it reverses every level including $0$; $\mathrm{NOT}_2 \equiv X_2$, and
$\mathrm{NOT}_d = X_d^{\dagger}K_d$. For $d=2^{m}$ these two gates have a particularly simple
reading, which is exactly their power-of-two definition in the code: $\mathrm{NOT}_d$ is the
bitwise complement, i.e.\ a transversal layer of \texttt{x} gates, and $K_d$ is the two's
complement $K_d = X_d\,\mathrm{NOT}_d$, emitted as a NOT gate followed by an X gate; at $d=2$ this
correctly collapses to $K_2 = X_2X_2 = I$.

\paragraph{SUMX / SUMXdg.} The qudit generalisation of CX. The repository implements it for an
arbitrary number $m_{\mathrm{c}} \geq 1$ of control qudits, of independent dimensions
$d_0,\dots,d_{m_{\mathrm{c}}-1}$, acting on a target of dimension $d_t$:
\begin{equation}
SUMX\,\lvert j_0\rangle\cdots\lvert j_{m_{\mathrm{c}}-1}\rangle\lvert k\rangle
= \lvert j_0\rangle\cdots\lvert j_{m_{\mathrm{c}}-1}\rangle\,
\big\lvert (k + j_0 + \cdots + j_{m_{\mathrm{c}}-1}) \bmod d_t\big\rangle .
\label{eq:sumx}
\end{equation}
For each fixed control assignment this is a cyclic shift of $\mathbb{Z}_{d_t}$, hence a permutation
and unitary. For $m_{\mathrm{c}}=1$ and $d_0=d_t=d$ it reduces to the single-control block-diagonal
form of~\cite{pudda2024} (blocks $I_d, X_d, X_d^2, \dots, X_d^{d-1}$ indexed by the control level).
SUMXdg subtracts instead of adding.

\paragraph{SUMP($\theta$).} The generalised controlled-phase gate, likewise generalised to
$m_{\mathrm{c}}$ controls:
\begin{equation}
SUMP(\theta)\,\lvert j_0\rangle\cdots\lvert j_{m_{\mathrm{c}}-1}\rangle\lvert k\rangle
= \omega_{d_t}^{\,(j_0+\cdots+j_{m_{\mathrm{c}}-1})\,k\,\theta/\pi}\,
\lvert j_0\rangle\cdots\lvert j_{m_{\mathrm{c}}-1}\rangle\lvert k\rangle ,
\label{eq:sump}
\end{equation}
reducing to~\cite[Eq.~(10)]{pudda2024} for $m_{\mathrm{c}}=1$ and equal dimensions. The gate is
diagonal with unit-modulus entries and therefore unitary for any combination of dimensions. Because
the phase factorises over the binary expansions of $j_a$ and $k$, its power-of-two definition is a
cascade of \texttt{cp} gates with angles weighted by the product of the two qubit significances.
The root of unity in Eq.~\eqref{eq:sump} is that of the \emph{target}: both the dense construction
and the \texttt{cp} cascade divide by the target dimension (\texttt{target\_dim} and
\texttt{target\_hilbert\_dim} respectively in the code), a distinction that is invisible for equal
dimensions but not for mixed ones. The power-of-two mixed-dimension rows of
Table~\ref{tab:verification-B} exercise exactly this: they are the only entries where the
\texttt{cp}/\texttt{mcx} cascade, rather than the dense fallback, is compared against a
mixed-dimension target.

\paragraph{SWAP.} $SWAP_d\lvert j\rangle\lvert k\rangle = \lvert k\rangle\lvert j\rangle$ for two
qudits of equal dimension $d$, built exactly as in~\cite{pudda2024}: $SUMX_d^\dagger$ (control
$q_1$, target $q_0$), then $SUMX_d$ (control $q_0$, target $q_1$), then $SUMX_d^\dagger$ (control
$q_1$, target $q_0$), then $K_d$ on $q_0$, mapping
$\lvert a,b\rangle \to \lvert a{-}b,b\rangle \to \lvert a{-}b,a\rangle \to \lvert{-}b,a\rangle \to
\lvert b,a\rangle$. For $d=4$ (power of two) the decomposed circuit shows this cascade:

{\footnotesize\begin{verbatim}      ┌────────┐        ┌────────┐┌────┐
qd_0: ┤0       ├───■────┤0       ├┤0   ├
      │ SUMXdg │   │    │ SUMXdg ││  K │
qd_1: ┤1       ├───■────┤1       ├┤1   ├
      └───┬────┘┌──┴───┐└───┬────┘└────┘
qd_2: ────■─────┤0     ├────■──────────
          │     │ SUMX │    │
qd_3: ────■─────┤1     ├────■──────────
                └──────┘
\end{verbatim}
}

Here qudits $0$ and $1$ each occupy two qubits ($d{=}4$), and the wire labels are again those of the
encoding qubits. The three SUMX-type boxes are \texttt{ControlledGate}s rendered as their
\texttt{base\_gate} (\texttt{Xdg} or \texttt{X}) with control dots on the other qudit's qubits; the
final $K$ box is a plain single-qudit \texttt{QuditGate} with no controls. Read left to right, the
cascade reproduces $SUMX^\dagger, SUMX, SUMX^\dagger, K$.

\paragraph{QFT.} Over $n$ qudits sharing a common dimension $d$,
\begin{equation}
QFT\lvert x\rangle = \frac{1}{\sqrt{d^{\,n}}}\sum_{y=0}^{d^{\,n}-1} \omega_{d^{n}}^{\,xy}\lvert y\rangle,
\qquad \omega_{d^{n}} = e^{i2\pi/d^{n}},\quad x = \sum_{i=0}^{n-1} x_i d^{\,i},
\label{eq:qft}
\end{equation}
with qudit $0$ holding the least-significant digit. (For $n=1$ this is $H_d^{\dagger}$, not $H_d$;
see Section~\ref{sec:notation}.) The decomposition applies, from most- to least-significant qudit,
an Hdg gate on the target followed by a $SUMP(\theta)$ from every less-significant qudit, with
\begin{equation}
\theta = \frac{\pi}{d^{\,s}}, \qquad s = \lvert q_c - q_t\rvert ,
\label{eq:qft-angle}
\end{equation}
and then restores digit order with a trailing cascade of SWAPs. Equation~\eqref{eq:qft-angle}
follows from Eq.~\eqref{eq:sump}: a $SUMP(\theta)$ contributes the phase
$\exp\!\big(i\,2j k\theta/d\big)$, while the QFT requires $\exp\!\big(i2\pi jk/d^{\,s+1}\big)$.
It is the corrected form of~\cite[Eq.~(14)]{pudda2024}, whose angle
$\pi/2^{\,d(q_c-q_t)}$ should read $\pi/d^{\,(q_c-q_t)}$. The construction is otherwise that
of~\cite{pudda2024}, generalised here from a single qudit pair to a register of $n$ qudits. For
$n=3$ qudits with $d=2$, the decomposed circuit is:

{\footnotesize
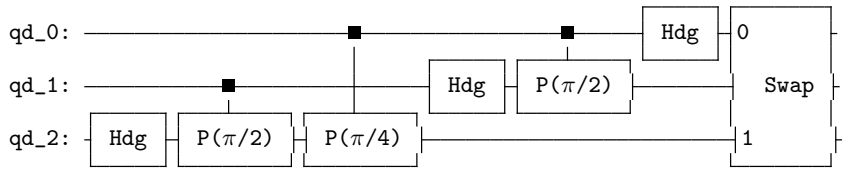
\begin{verbatim}                                                  ┌─────┐┌───────┐
qd_0: ─────────────────────■────────────────■─────┤ Hdg ├┤0      ├
                           │     ┌─────┐┌───┴────┐└─────┘│       │
qd_1: ───────────■─────────┼─────┤ Hdg ├┤ P(π/2) ├───────┤  Swap ├
      ┌─────┐┌───┴────┐┌───┴────┐└─────┘└────────┘       │       │
qd_2: ┤ Hdg ├┤ P(π/2) ├┤ P(π/4) ├────────────────────────┤1      ├
      └─────┘└────────┘└────────┘                        └───────┘
\end{verbatim}
}

with $\theta=\pi/2$ at distance $s=1$ and $\theta=\pi/4$ at distance $s=2$, matching the
Hdg-then-controlled-phase-cascade-then-swap structure of~\cite[Fig.~2]{pudda2024}. When $d$ is not a
power of two (e.g.\ $d=3$), the same \texttt{.qft()} call produces a circuit with the same encoded
operator but no visible structure: it decomposes to a single dense \texttt{unitary} instruction
rather than this cascade, exactly as for the qutrit SWAP of Section~\ref{sec:paths}. The cost of
that choice is analysed in Section~\ref{sec:limitations}.

\section{Verification: the QFT Unitary Against the Discrete Fourier Transform}
\label{sec:verification}

The construction of~\cite{pudda2024} was verified only for $d = 2^{m}$ with $m\le4$, i.e.\ for
qudits perfectly simulable by $m$ qubits; the simulation could not be pushed further for memory
reasons, and non-power-of-two dimensions were out of reach altogether. Here we run two distinct
checks, and we are explicit about what each one does and does not establish, because they are often
conflated.

\begin{description}
\item[Check A (definition vs.\ DFT).] Does the dense matrix returned by
\texttt{\_build\_unitary()} reproduce the discrete Fourier transform on the logical subspace? This
validates Eq.~\eqref{eq:qft}, the index map of Eq.~\eqref{eq:isometry} and the sign convention. It
does \emph{not} exercise \texttt{\_define()} at all, and therefore cannot detect an error in any
decomposition; whether the identity-padding invariant itself holds is a separate question, checked
below.
\item[Check B (decomposition vs.\ definition).] Does the circuit emitted by \texttt{\_define()}
implement the same operator as \texttt{\_build\_unitary()}? This is the check that corresponds to
what~\cite{pudda2024} did by simulation, and it is the one that would catch, for example, a wrong
\texttt{inverse}/\texttt{do\_swaps} flag in the \texttt{synth\_qft\_full} call or a wrong angle in
Eq.~\eqref{eq:qft-angle}. For non-power-of-two $d$ the fallback path embeds
\texttt{\_build\_unitary()} verbatim, so Check B is trivially satisfied there and is informative
only on the power-of-two path.
\end{description}

Note that what makes both checks available for arbitrary $d$ is the \texttt{\_\_array\_\_}
protocol of Section~\ref{sec:hierarchy}, which exposes the dense encoded unitary independently of
which decomposition path \texttt{\_define()} takes. All numbers in this section were obtained with
Python~3.12.3, Qiskit~2.5.2, NumPy~2.5.2 and SciPy~1.18.0.

\subsection{Check A}

The encoded $D \times D$ matrix is restricted to the valid $d^{\,n}$-dimensional subspace using
\texttt{\_encoded\_index} and compared against \texttt{scipy.linalg.dft(d**n, scale='sqrtn')}.
Because the gate uses the $\omega_{d^n} = e^{+i2\pi/d^{n}}$ convention while SciPy's \texttt{dft}
uses $e^{-i2\pi/L}$, the reference is conjugated before comparison. We additionally assert the
invariant of Eq.~\eqref{eq:invariant} itself, i.e.\ that the complement block is the identity and
that there is no valid/invalid mixing; this is not implied by Check A and is never tested otherwise.

\begin{lstlisting}[style=python]
import numpy as np
from scipy.linalg import dft
from qiskit_qudits.gates import QuditQFTGate
from qiskit_qudits.utils.encoding import _encoded_index

def valid_indices(dims):
    return [_encoded_index(dims, i) for i in range(int(np.prod(dims)))]

def check_A(n, d):
    """Logical block of the dense unitary == DFT^dagger."""
    gate = QuditQFTGate(n, d)
    unitary = np.array(gate)
    dims = [d] * n
    idx = valid_indices(dims)
    sub = unitary[np.ix_(idx, idx)]
    reference = dft(d ** n, scale='sqrtn').conj()
    return np.max(np.abs(sub - reference))          # report the deviation, not a bool

def check_padding(gate, dims):
    """Complement block == identity and no valid/invalid mixing."""
    U = np.array(gate)
    D = U.shape[0]
    idx = valid_indices(dims)
    rest = sorted(set(range(D)) - set(idx))
    err_id  = np.max(np.abs(U[np.ix_(rest, rest)] - np.eye(len(rest)))) if rest else 0.0
    err_off = max(np.max(np.abs(U[np.ix_(idx, rest)])),
                  np.max(np.abs(U[np.ix_(rest, idx)]))) if rest else 0.0
    return err_id, err_off
\end{lstlisting}

We separately confirmed that $U=\texttt{numpy.array(gate)}$ is unitary to within
$\lVert U^{\dagger}U-I_D\rVert_{\max}<3\times10^{-15}$ for all four configurations of
Table~\ref{tab:verification-A}; together with \texttt{err\_id}$=0$ this already implies
\texttt{err\_off}$=0$ (a unitary matrix whose valid columns have unit norm inside the logical block
cannot have any support outside it). We nonetheless report \texttt{err\_off} directly, since it
makes no unstated assumption about $U$.

Note that \texttt{np.allclose} keeps its default \texttt{rtol=1e-5} unless it is set to zero, so a
Boolean comparison with \texttt{atol=1e-10} alone would in fact be testing at a tolerance of about
$10^{-5}/\sqrt{d^{\,n}}$. We therefore report the achieved deviation directly.

\begin{table}[h]
\centering
\begin{tabular}{ccccccccc}
\toprule
$n$ & $d$ & $m$ & power of two? & $N$ & $D=2^{N}$ & $d^{\,n}$ & $\max\lvert\Delta\rvert$ (Check A) & padding \\
\midrule
3 & 2 & 1 & yes & 3 & 8  & 8  & $8.6\times10^{-16}$ & $0$ (exact) \\
2 & 4 & 2 & yes & 4 & 16 & 16 & $2.8\times10^{-15}$ & $0$ (exact) \\
2 & 3 & 2 & no  & 4 & 16 & 9  & $1.2\times10^{-15}$ & $0$ (exact) \\
2 & 5 & 3 & no  & 6 & 64 & 25 & $4.3\times10^{-15}$ & $0$ (exact) \\
\bottomrule
\end{tabular}
\caption{Check A: logical block of \texttt{numpy.array(QuditQFTGate(n, d))} against
$\mathrm{DFT}^{\dagger}$, for two power-of-two and two non-power-of-two dimensions. The last column
reports $\max(\text{err\_id},\text{err\_off})$ from \texttt{check\_padding}, i.e.\ the violation of
Eq.~\eqref{eq:invariant}; it vanishes \emph{exactly}, not merely to numerical precision, because
the padding entries are written as exact $0$s and $1$s and never touched again.}
\label{tab:verification-A}
\end{table}

\subsection{Check B}

Check B compares the operator of the emitted definition with the dense matrix:

\begin{lstlisting}[style=python]
from qiskit.quantum_info import Operator

def check_B(n, d):
    """Operator of _define() == dense unitary of _build_unitary()."""
    gate = QuditQFTGate(n, d)
    return np.max(np.abs(Operator(gate.definition).data - np.array(gate)))
\end{lstlisting}

For the four configurations of Table~\ref{tab:verification-A} the deviations are
$1.2\times10^{-15}$ ($n{=}3$, $d{=}2$), $2.0\times10^{-15}$ ($n{=}2$, $d{=}4$), and exactly $0$ for
the two non-power-of-two rows, as expected: there the fallback path embeds
\texttt{\_build\_unitary()} verbatim. Beyond the QFT, Check B and the padding check were run over
the \emph{entire} gate set of Table~\ref{tab:gateset} for every dimension $2\le d\le 8$, plus five
mixed-dimension operand tuples (two using only power-of-two operands, to exercise the
\texttt{cp}/\texttt{mcx} cascade, and three using at least one non-power-of-two operand, to
exercise the dense fallback) $138$ gate instances in total. Table~\ref{tab:verification-B}
reports the worst deviation per gate; the padding check of Eq.~\eqref{eq:invariant} is satisfied
exactly (deviation $0$) for every instance.

\begin{table}[h]
\centering
\footnotesize
\begin{tabularx}{\linewidth}{@{} p{4.2cm} l >{\raggedright\arraybackslash}X @{}}
\toprule
Gate(s) & $\max$ Check B deviation over $2\le d\le 8$ & Note \\
\midrule
I, X / Xdg, S / Sdg, K, NOT, SWAP & $0$ (exact) & permutations/pinned phases exact \\
Z / Zdg & $1.1\times10^{-16}$ & \\
P($0.7$) & $1.6\times10^{-16}$ & \\
T / Tdg & $2.2\times10^{-16}$ & \\
H / Hdg & $5.7\times10^{-16}$ & exercises \texttt{synth\_qft\_full} \\
SUMX / SUMXdg & $3.3\times10^{-16}$ & \\
SUMP($0.7$) & $5.7\times10^{-16}$ & \\
QFT ($n{=}2$) & $5.0\times10^{-15}$ & worst case, at $d=8$ \\
\midrule
SUMX $(2,4)\!\to\!8$ & $4.5\times10^{-16}$ & mixed, power-of-two (named-gate cascade) \\
SUMP($0.7$) $(2,4)\!\to\!8$ & $4.0\times10^{-16}$ & mixed, power-of-two (named-gate cascade) \\
SUMX $(3)\!\to\!5$,\newline SUMX $(2,5)\!\to\!3$,\newline SUMP($0.7$) $(3,2)\!\to\!4$ & $0$ (exact) & mixed, at least one non-power-of-two operand (fallback path) \\
\bottomrule
\end{tabularx}
\caption{Check B over the whole gate set: maximum of
$\lvert\texttt{Operator(gate.definition)} - \texttt{numpy.array(gate)}\rvert$ across
$d\in\{2,\dots,8\}$ per gate, plus mixed-dimension controlled gates (control dims $\to$ target
dim). Exact zeros arise on the non-power-of-two path (verbatim embedding) and for power-of-two
permutation gates, whose cascades are exact in integer arithmetic; the power-of-two mixed-dimension
rows are the only entries that exercise the \texttt{cp}/\texttt{mcx} cascade with unequal operand
dimensions, and confirm the target-dimension convention of Eq.~\eqref{eq:sump}. The padding check
is exactly zero for all $138$ instances.}
\label{tab:verification-B}
\end{table}

Together, Check A on the non-power-of-two rows extends the numerical validation
of~\cite{pudda2024} to dimensions its qubit-simulation approach could not reach, while Check B on
the power-of-two rows reproduces, and on the remaining gates extends, the property that
paper actually verified.

\section{Measurement and Classical Decoding}
\label{sec:measurement}

Measurement is not a unitary gate, so it is handled by \texttt{QuditMeasure}, one of a family of
\texttt{QuditDirective} classes (\texttt{circuit/directives.py}) that subclass Qiskit's
\texttt{Instruction} directly: instead of a \texttt{\_define()}, every such object implements
\texttt{apply(circuit, qudits, clbytes)}, which is dispatched by
\texttt{QuditQuantumCircuit.\_apply\_operation} and emits primitive Qiskit operations onto the
encoded circuit. \texttt{QuditMeasure} measures encoding qubit $j$ of each target qudit straight
into clbit $j$ of its paired \texttt{ClByte}. A note on the name: in Qiskit, \emph{directive}
(\texttt{Instruction.\_directive}) means specifically a compiler hint that is not a real operation,
such as \texttt{barrier}. The library respects that flag at the instance level (only
\texttt{QuditBarrier} sets \texttt{\_directive=True}, so measurements, resets and initialisations
are counted by \texttt{size()}/\texttt{depth()} exactly as in Qiskit) but the base-class
\emph{name} is broader than Qiskit's usage: a \texttt{QuditDirective} is any non-unitary qudit
operation expanded through \texttt{apply()}, not necessarily a compiler hint.

Decoding is a purely classical post-processing step in \texttt{utils/encoding.py}.
\texttt{decode\_bitstring}/\texttt{decode\_counts} split a backend counts key into one chunk per
clbyte, of width $m_i=\lceil\log_2 d_i\rceil$, so chunk widths differ in a mixed-dimension
register, reading each chunk MSB-first and accounting for Qiskit's convention that clbit $i$ sits
at string position $\mathrm{len}-1-i$; each chunk is then reinterpreted as \texttt{int(chunk, 2)}.
Chunk widths are always taken per clbyte from that clbyte's own dimension
(\texttt{clbyte\_widths}), never assumed uniform, and whitespace, Qiskit's separator between
classical registers, is stripped before splitting, so multi-register counts keys decode
transparently. Both behaviours were exercised on a mixed $(2,3,5)$ register with chunk widths
$(1,2,3)$: the key \texttt{'100101'} and its spaced variant \texttt{'100 101'} both decode to
levels $(1,2,4)$.

Because the encoding fills the $2^{m}$-dimensional space exactly only for power-of-two $d$, a
measured bitstring can decode to a value $\ge d$ that the qudit was never supposed to reach, a
leakage event. Under the invariant of Section~\ref{sec:invariant}, no gate can create such an
outcome, so in an ideal, noiseless simulation of a circuit built entirely through the qudit API
leakage cannot occur; it becomes possible only under a noise model or on hardware, when
\texttt{QuditInitializeLevels}/\texttt{QuditStatePreparation} are given out-of-range input, or when
the user has written directly to \texttt{.circuit} (Section~\ref{sec:qqc}). Detecting it is
therefore a useful integrity check on the simulation, not a routine occurrence. An
\texttt{InvalidPolicy} argument (\texttt{'keep' | 'drop' | 'raise'}) governs what
\texttt{decode\_bitstring}/\texttt{decode\_counts} do when it happens, making leakage detection an
explicit, opt-in part of the decoding API rather than a silent failure mode. Note that
\texttt{'drop'} discards the offending shots, so the returned counts no longer sum to the number of
shots and the resulting distribution is conditioned on the absence of leakage; users comparing
distributions across noise levels should prefer \texttt{'keep'} or \texttt{'raise'} and report the
leakage fraction separately. \texttt{format\_levels} then renders decoded levels in Qiskit's own
reversed, space-separated display convention.

Four further non-unitary qudit instructions round out the state-preparation API:
\texttt{QuditBarrier}; \texttt{QuditReset} (which resets every encoding qubit of a qudit, including
any leaked ones); \texttt{QuditInitializeLevels} (basis-state preparation from a bitstring label,
via \texttt{QuantumCircuit.initialize}); and \texttt{QuditStatePreparation} (arbitrary-amplitude
preparation over $L=\prod_i d_i$ logical amplitudes, embedded into and projected back from the
$D=2^{N}$-dimensional encoded space by \texttt{embed\_state}/\texttt{project\_state}). Two practical
caveats: \texttt{QuantumCircuit.initialize} contains a reset and is therefore not accepted
unmodified by every backend or primitive, which qualifies the ``runs anywhere'' claim of
Section~\ref{sec:qqc}; and \texttt{project\_state} is norm-preserving only for states supported
inside $\Pi$; when the input carries amplitude outside the qudit subspace beyond an absolute
tolerance of $10^{-8}$, it neither renormalises silently nor warns: it raises a
\texttt{ValueError} reporting the leaked norm.

\section{Limitations and Design Trade-offs}
\label{sec:limitations}

\paragraph{Supported dimensions.} The mixins cap dimensions at $2\le d\le16$ and multi-gates at $8$
qudits; both caps are enforced eagerly by parameter validation (a \texttt{ValueError} at
construction time). Within that window, non-power-of-two dimensions are handled exactly; outside it
the library raises. The dimension cap corresponds to at most $m=4$ encoding qubits per qudit, and
together with the qudit cap bounds the nominal instruction width at $N\le32$; the caps are not
memory-derived, and the binding constraint in practice is the dense fallback: a $D\times D$
\texttt{complex128} matrix costs $16\cdot4^{N}$ bytes ($1$\,GiB at $N=13$, $4$\,GiB at $N=14$), so
the effective joint limit on $(d,n)$ for the dense path is $\sum_i m_i \lesssim 13$--$14$ on
commodity hardware. Exceeding it fails only lazily: e.g.\ a gate on eight $16$-level qudits
constructs successfully, and its \texttt{\_build\_unitary()} then dies with a NumPy
\texttt{MemoryError} (an allocation of $16\cdot4^{32}\approx2.95\times10^{20}$ bytes, i.e.\
$256$\,EiB, at $N=32$) rather than a library-level error; an eager memory check is future work.
Claims of ``arbitrary $d$'' elsewhere in the literature should be read against this cap.

\paragraph{Cost of the dense fallback.} For non-power-of-two operands, \texttt{\_define()} emits one
$D\times D$ unitary for the \emph{whole} instruction; this holds for the QFT too, which is
collapsed into a single register-wide \texttt{unitary} rather than a cascade of dense qudit gates.
For a two-qudit gate this is harmless ($D\le 256$), but for an $n$-qudit QFT it means building and
synthesising a $2^{n\lceil\log_2 d\rceil}$-dimensional matrix: a $d=3$ QFT over six qutrits already
requires a $4096\times4096$ unitary (built in $0.10$\,s; its quantum Shannon decomposition costs
$O(4^{12})$ two-qubit gates and was not attempted). Table~\ref{tab:resources} quantifies the
effect at $n=3$: the $d=3$ QFT ($N=6$, encoded dimension $D=64$ around a $27$-dimensional logical
block) synthesises to $1{,}783$ CX gates against $56$ for its power-of-two neighbour $d=4$, and the
$d=5$ QFT ($N=9$, $D=512$ around a $125$-dimensional logical block) to $119{,}383$ CX against $156$
for $d=8$. This is a property of the current implementation, not of the construction: the QFT
cascade of Section~\ref{sec:gateset} is dimension-agnostic, so only its one- and two-qudit
constituents need dense synthesis, which would reduce the cost to polynomial in $n$. The same
observation applies to multi-control SUMX/SUMP. Similarly, the permutation gates (X, K, NOT, SUMX,
SWAP) are reversible classical functions for every $d$ and admit multi-controlled-X synthesis
directly, without going through a dense matrix.

\paragraph{Redundant cost on the power-of-two path.} The named-gate cascades of
Section~\ref{sec:gateset} are themselves not optimal. Two checks make this concrete. First, because
$d^{n}=2^{N}$ whenever $d$ is a power of two, the $n$-qudit QFT of Eq.~\eqref{eq:qft} is, as an
operator, exactly Qiskit's flat $N$-qubit QFT (\texttt{synth\_qft\_full(N)}); the two agree to
$1.5\times10^{-14}$ or better for $n=3$ and $d\in\{2,4,8\}$. Transpiled to the same basis, the flat
QFT is markedly cheaper than the current per-qudit cascade of Table~\ref{tab:resources}: $6$ vs.\
$9$ CX at $d=2$, $30$ vs.\ $56$ at $d=4$, and $72$ vs.\ $156$ at $d=8$. Second, for power-of-two $d$
the qudit SWAP is, on the logical subspace, equivalent to a transversal layer of $m$ ordinary qubit
\texttt{swap} gates rather than the four-gate SUMX-based cascade of Table~\ref{tab:gateset};
transpiled in isolation this permutation is absorbed entirely into the output layout (zero
two-qubit gates measured, versus $23$ CX at $d=4$ and $84$ CX at $d=8$ for the current cascade),
and even when the permutation must instead be materialised explicitly it costs at most $3m$ CX.
Both observations quantify the future-work item of Section~\ref{sec:conclusion} on
dimension-agnostic, reversible-logic synthesis of the one- and two-qudit constituents rather than
per-instruction cascades or dense matrices.

\paragraph{Resource data.} Table~\ref{tab:resources} reports, for representative gates and
dimensions, the encoded qubit count, the CX count and depth after
\texttt{transpile(..., basis\_gates=['cx','u'], optimization\_level=2)} on Qiskit~2.5.2, and the
wall-clock times to build the dense matrix and to transpile (single run, indicative only). The
pattern substantiates the ``transpiler-friendly'' claim for power-of-two dimensions and makes the
cost of the dense fallback explicit: within each gate, the non-power-of-two rows ($d=3,5$) are more
expensive than their power-of-two neighbours ($d=4,8$) of the same encoded width, by a factor
ranging from about unity for the narrowest gates (e.g.\ H at $d=3$ vs.\ $d=4$: $3$ vs.\ $3$ CX) up
to nearly three orders of magnitude for the widest ones (QFT at $d=5$ vs.\ $d=8$: $119{,}383$ vs.\
$156$ CX, a factor of approximately $765$).

\begin{table}[h]
\centering
\footnotesize
\begin{tabular}{llrrrrr}
\toprule
Gate & $d$ & qubits $N$ & CX & depth & $t_{\mathrm{build}}$ [ms] & $t_{\mathrm{transpile}}$ [ms] \\
\midrule
X    & 2 & 1 & 0 & 1 & 0.04 & 5.1 \\
X    & 3 & 2 & 2 & 5 & 0.09 & 11.1 \\
X    & 4 & 2 & 1 & 2 & 0.06 & 5.6 \\
X    & 5 & 3 & 16 & 32 & 0.10 & 30.4 \\
X    & 8 & 3 & 5 & 11 & 0.05 & 16.5 \\
\midrule
H    & 2 & 1 & 0 & 1 & 0.09 & 5.6 \\
H    & 3 & 2 & 3 & 7 & 0.14 & 10.0 \\
H    & 4 & 2 & 3 & 7 & 0.09 & 10.6 \\
H    & 5 & 3 & 19 & 39 & 0.85 & 24.0 \\
H    & 8 & 3 & 6 & 13 & 0.11 & 13.4 \\
\midrule
SUMX & 2 & 2 & 1 & 1 & 0.09 & 6.9 \\
SUMX & 3 & 4 & 52 & 101 & 0.16 & 20.8 \\
SUMX & 4 & 4 & 6 & 10 & 0.13 & 14.0 \\
SUMX & 5 & 6 & 1002 & 1980 & 0.18 & 96.4 \\
SUMX & 8 & 6 & 23 & 38 & 0.32 & 8.9 \\
\midrule
SWAP & 2 & 2 & 3 & 3 & 0.07 & 12.5 \\
SWAP & 3 & 4 & 91 & 179 & 0.04 & 16.8 \\
SWAP & 4 & 4 & 23 & 34 & 0.04 & 15.8 \\
SWAP & 5 & 6 & 1755 & 3468 & 0.07 & 116.6 \\
SWAP & 8 & 6 & 84 & 136 & 0.05 & 16.7 \\
\midrule
QFT ($n{=}3$) & 2 & 3 & 9 & 16 & 0.23 & 19.3 \\
QFT ($n{=}3$) & 3 & 6 & 1783 & 3525 & 0.18 & 140.2 \\
QFT ($n{=}3$) & 4 & 6 & 56 & 77 & 0.57 & 19.3 \\
QFT ($n{=}3$) & 5 & 9 & 119383 & 236036 & 2.09 & 8269.6 \\
QFT ($n{=}3$) & 8 & 9 & 156 & 204 & 12.86 & 16.1 \\
\bottomrule
\end{tabular}
\caption{Resources after transpilation to the generic basis \texttt{['cx','u']} at
\texttt{optimization\_level=2} (Qiskit~2.5.2). Non-power-of-two rows ($d=3,5$) go through the dense
fallback; power-of-two rows through the named-gate cascades. Times are single-run wall-clock
measurements and indicative only.}
\label{tab:resources}
\end{table}

\paragraph{Encoding overhead.} Storing a $d$-level system in $2^{\lceil\log_2 d\rceil}$ levels costs
a factor $\big(2^{\lceil\log_2 d\rceil}/d\big)^{n}$ in statevector memory relative to a native-qudit
simulator: $1.33^{n}$ for $d=3$, $1.6^{n}$ for $d=5$, $1.78^{n}$ for $d=9$; the worst case in the
supported range, since the factor is always strictly below $2$. The trade-off is
deliberate, it is what buys compatibility with the entire Qiskit stack, but for pure
simulation studies at large $n$ a native-qudit
simulator~\cite{kushnarev2025,lambert2026,chatterjee2023,mato2024} will be more economical.

\paragraph{Leakage semantics.} As discussed in Section~\ref{sec:invariant}, identity padding freezes
leaked amplitude instead of propagating it, and does not commute with tensoring. Circuit
equivalence must therefore be assessed on $\Pi U\Pi$.

\section{Availability and Reproducibility}
\label{sec:availability}

The numerical results of Sections~\ref{sec:verification} and~\ref{sec:limitations} were produced
with Python~3.12.3, Qiskit~2.5.2, NumPy~2.5.2 and SciPy~1.18.0. Repository is available at
\url{https://github.com/Frank995/qiskit-qudits}

\section{Conclusion}
\label{sec:conclusion}

\texttt{qiskit-qudits} turns the generalised qudit gate set and QFT construction
of~\cite{pudda2024} into a Qiskit extension built on Qiskit's own \texttt{Gate},
\texttt{ControlledGate} and \texttt{Instruction} classes, so that qudit circuits transpile and run
like any other Qiskit circuit. Its central design choice (acting as the identity on every
out-of-subspace basis state (Equation~\eqref{eq:invariant}), then dispatching between a
transpiler-friendly cascade for power-of-two dimensions and exact dense unitary synthesis otherwise)
lets it realise qudit dimensions $2\le d\le16$ exactly, and not only the powers of two that the
original theoretical work could verify by qubit simulation. Section~\ref{sec:verification} confirms
this numerically: the QFT against the discrete Fourier transform, every gate's emitted
decomposition against its dense unitary across the whole gate set ($138$ instances, worst-case
deviation $5\times10^{-15}$), and the padding invariant the encoding relies on, which holds
exactly; transpile round-trips of the encoded circuits are operator-equivalent at all four
optimisation levels. The measurement and decoding layer makes leakage out of the qudit subspace an
explicit, checkable property of a simulation rather than a silent source of error. The main
directions for future work follow from Section~\ref{sec:limitations}: applying the dense fallback
at the smallest possible operand size rather than per instruction, using reversible-logic synthesis
for the permutation gates, recognising the power-of-two identities of Section~\ref{sec:limitations}
(the flat $N$-qubit QFT, the transversal SWAP) instead of the current per-instruction cascades, and
exposing the unphysical subspace to the transpiler as a don't-care region.

\section*{Acknowledgements}
I would like to thank Luca Crippa for spinning up this project back in the day.

\printbibliography

\end{document}